\documentclass[showpacs,aps]{revtex4}
\usepackage{amsmath}
\usepackage{graphicx}
\begin{document}

\title{\bf  Poisson type relativistic perfect fluid spheres}

% from potential-density pairs

%formato prd

\author{Gonzalo Garc\'{\i}a-Reyes}
\email[e-mail: ]{ggarcia@utp.edu.co}
\affiliation{Departamento de F\'{\i}sica, Universidad Tecnol\'ogica de Pereira,
 A. A. 97, Pereira, Colombia}

 \begin{abstract}

 Static spherically  symmetric  solutions of the
 Einstein's field equations  in isotropic coordinates  representing  perfect  fluid  matter distributions
 from  Newtonian potential-density pairs are investigated.
The   approach 
is illustrated  with  three  simple examples based on the
potential-density pairs corresponding to a  harmonic oscillator (homogeneous sphere), the well-known Plummer model   and a  massive spherical dark matter halo model with a logarithmic potential.  
Moreover,  the geodesic circular  motion of test particles  around  such structures is studied. 
The stability of the  orbits against radial perturbations is also analyzed  using an extension of the Rayleigh criteria of stability of a fluid in rest in a gravitational field.  The  models considered  satisfy  all the energy conditions.  
\end{abstract}

%Keywords:  General relativity; Exact solutions; Spherical symmetric; Perfect fluid

\maketitle

\section{Introduction}

Matter distributions   with  spherical symmetry have     played an important role  in astrophysics   as models of  dwarf spheroidal galaxies 
\cite{Wilkinson},   bulges of disc galaxies \cite{Bajkova1,Bajkova2},  galactic nuclei  \cite{Dehnen,Tremaine},  globular clusters \cite{Plummer},  clusters of galaxies and dark matter haloes \cite{Hernquist,NFW}.  In relativistic astrophysics, such
matter configurations have been used as models of 
neutron stars, highly dense stars,
gravastars, dark energy stars, quark stars, galactic nuclei and certain
star clusters where relativistic effects are expected to be important.  
 In relation to  static spherically symmetric fields  several exact solutions have been obtained  over the years \cite{Delgaty,Kramer}. 
In particular,  anisotropic relativistic models of spherically symmetric matter distributions  from various Newtonian  potential-density pairs in isotropic coordinates  were investigated in Refs. \cite{Let-shell1,Let-shell2} for  Schwarzschild type  space-times, also used in     \cite{Nguyen} for the same space-time,  and in  Ref. \cite{GGR-shell} for Majumdar-Papapetrou type fields.  This same approach and spacetime was also  used in  \cite{GGR-K} in the  construction of three-dimensional anisotropic axisymmetric sources. On the other hand, perfect fluid matter distributions  as a source of the gravitational
field have also been widely studied in general
relativity, mainly to describe models of stars, galaxies,
and universes  \cite{Tolman, Zeldovich}.

In this work, we investigate  the construction of  
perfect fluid sources   for static spherically  symmetric fields from a Newtonian potential-density pair using isotropic coordinates.  These spacetimes contain   two unknown metric functions. By substitution,  one of the Einstein's field equations reduces to  a nonlinear Poisson
type equation. This
equation is  solved for one of the metric functions  based on the
assumption that in the Newtonian limit we should get a solution  of the  Poisson's equation of  Newtonian gravity  and based on a special form of the energy density.
The other metric function is obtained solving 
the condition of  pressure    isotropy for a perfect fluid matter distribution.
The method is illustrated  with three simple examples based  on the potential-density pairs of  a homogeneous sphere, the well-known  Plummer  model and a logarithmic potential.   These potential-density pairs have been used  by some authors  as models of dark matter haloes \cite{Wilkinson, Bajkova2,Binney} and  although such structures  are essentially  Newtonian,
their relativistic description  can be  interesting  not only from the merely theoretical point of view but also  in the analysis  of their   interaction 
with large scale cosmic evolution and  relativistic effects such as gravitational lenses or gravitational waves \cite{Matos}.

The paper  is structured as follows.  In Sect. II we present the method used for the construction of 
static spherically  symmetric perfect fluid exact solutions 
from  a given solution of  Poisson's  equation   using isotropic coordinates.   We
also analyse the geodesic motion of  test particles around the spherical distributions of matter 
and the stability of the orbits against radial perturbations using an extension of the
Rayleigh criteria of stability of a fluid in rest in a gravitational field.  In Sects. III-V  the method  is illustrated  with three simple examples using as seed potential-density pairs those  of a homogeneous sphere, the Plummer sphere  and dark matter halo model with a logarithmic potential. Finally, in Sect. VI we summarize and discuss the results obtained.

%---------------------------------------------------------------------
%---------------------------------------------------------------------

\section{ Relativistic perfect fluid spheres } 
The line element for   a static spherically symmetric spacetime  in isotropic coordinates  is given by \cite{Kramer}
\begin{equation}
ds^2 =  e^{2\nu} c^2 dt^2 - e^{2\lambda}  (dr^2 + r^2d\Omega^2),  \label{eq:met}
\end{equation}
%\left(1- \frac {\phi(r)}{ 2 c^2}  \right)^4
where   $d\Omega^2 = d\theta^2 +  \sin ^2 \theta  d\varphi ^2$, and  $\nu$ and  $\lambda$  are functions of only $r$. We use coordinates $(x^0, x^1,x^2,x^3)=(ct,r,\theta,\varphi)$.  Einstein's gravitational  field equations $G_{ab}= (8 \pi G /c^4)  T_{ab} $ reduce to
\begin{subequations}\begin{eqnarray}
T^t_{\ t} & = &  \frac{c^2}{4 \pi G} e^{ -2\lambda} \left [\nabla^2 \lambda +  \frac1 2 \nabla \lambda \cdot \nabla \lambda \right ], \\
%  \frac{1}{4 \pi G} e^{ -2\lambda} \left [ \lambda '' + \frac{2\lambda' }{r} + \frac{(\lambda')^2 } {2} \right ],   \\
&  &  \nonumber \\
 T^r_{\ r} & = & \frac{c^4}{8 \pi G} e^{ -2\lambda} \left [ (\lambda')^2 + 2\lambda' \nu' + \frac {2} {r} (\lambda' + \nu' )  \right ]  ,  \label{eq: p1}  \\
& &  \nonumber  \\
T^\theta_{ \ \theta} & = & T^\varphi_{ \ \varphi}  = \frac{c^4}{8 \pi G} e^{ -2\lambda} \left [
\lambda '' + \nu '' + (\nu')^2 + \frac {1} {r} (\lambda' + \nu' ) \right] ,    \label{eq: p2} 
\end{eqnarray} \end{subequations} 
where primes indicate differentiation with respect to $r$.

In terms of the orthonormal  tetrad   (comoving observer) ${{\rm e}_{ (a)}}^b = \{
U^b , X^b,Y^b,  Z^b \}$, where
\begin{subequations}\begin{eqnarray}
U^a & = & \frac{1}{\sqrt{g_{00}}} \delta ^a_{0} =
  e^{-\nu}    \delta ^a_{0}  , \quad  \quad \quad 
X^a  = \frac{1}{\sqrt{-g_{11}}}  \delta ^a_{1} = e^{-\lambda} \delta ^a_{1} , \quad   \label{eq:tetrad1} \\
Y^a  &= & \frac{1}{\sqrt{-g_{22}}}  \delta ^a_{2}  =   \frac 1 r e^{-\lambda}   \delta ^a_{2} ,  \	 \quad
Z^a = \frac{1}{\sqrt{-g_{33}}}  \delta ^a_{3} = \frac{1}{r \sin  \theta}  e^{-\lambda}     \delta ^a_{3}  . \label{eq:tetrad2}
\end{eqnarray}\end{subequations}
the energy density is  $\rho=T^0_{\ 0}/c^2$  and
the principal stresses  $p_i=-T^i_{\ i}$.  $U^a$ is the four-velocity of the matter distributions  defined as  the timelike vector  of the orthonormal  tetrad. 

By making   $e^{2 \lambda} = \left( 1- \frac{\phi}{2c^2}    \right)  ^{4}$, 
we get for the energy density  the following    nonlinear  Poisson type equation
\begin{equation}
\nabla ^2 \phi = 4 \pi G  \rho  \left( 1- \frac{\phi}{2} \right)^{5} . \label{eq: nonlinear}
\end{equation}
For  a  given physical energy density  profile $\rho$,
the metric function $\phi$ can be obtained 
by resolving this equation.  A physically reasonable way  to choose $\rho$ is  
by requiring  that in the Newtonian limit it   reduces to its  Newtonian value $\rho_N$.  
A  simple particular form of   $\rho$ which satisfies such condition  is
\begin{equation}
\rho  =   \frac{\rho_0 }{   \left( 1- \frac{\phi}{2 } \right)^{5} }.  \label{eq: rho}
\end{equation}
Replacing   this expression in (\ref{eq: nonlinear}) one finds in this case that the pair  $(\phi,\rho_0 )$ is a  solution of the Poisson's equation. Hence  $(\phi,\rho_0 )= (\Phi,\rho_N )$ for any physical system.  Therefore, 
\begin{equation}
\rho  =   \frac{\rho_N }{   \left( 1- \frac{\Phi}{2 } \right)^{5} }.  \label{eq: rho}
\end{equation}

To obtain the  other metric function $\nu$  an  additional assumption  must be imposed.
For a perfect fluid  source the momentum-energy tensor has the form \cite{LAND2,LAND6,Taub,Synge,Misner,Hawking,Schutz}
\begin{equation}
T^{ab} = (\rho c^2 + p)U^aU^b  - p g^{ab},
\end{equation}
and such assumption is  the condition of  pressure    isotropy
\begin{equation}
\lambda '' + \nu '' + (\nu')^2 - (\lambda')^2 - 2\lambda' \nu'
-  \frac {1} {r} (\lambda' + \nu' ) = 0,  \label{eq: isotropy} 
\end{equation}
which is a Riccati equation in either $\lambda'$ or $\nu'$. This condition is obtained  by equating the field equations  (\ref{eq: p1}) and (\ref{eq: p2}).  
It can also be cast as \cite{Qvist}
\begin{equation}
L {\cal G}_{,xx} = 2  {\cal G} L_{,xx}, \ \ L \equiv e^{- \lambda}, \ \  {\cal G} \equiv L e^{\nu}, \ \ x \equiv r^2.  \label{eq:iso}
\end{equation}

Thus,  this  approach  allows  to construct  different static spherically  symmetric  solutions in isotropic coordinates  representing  perfect  fluid  matter distributions from a given solution of  Poisson's equation. 
In addition, in order to have  physically meaningful  sources   the parameters of the solutions
must be chosen so that   the energy conditions are satisfied. For a perfect fluid,  the weak energy condition reads
$\rho\geq 0$,  whereas   the dominant energy condition states that 
$|\rho|  \geq |p|$.   The strong energy
condition requires that 
$\rho_{eff} = \rho+ 3p \geq 0$,  where $\rho_{eff}$
is the ``effective Newtonian density''. 

An important quantity  related to the circular  motion of test particles  around matter distributions is the circular speed
(rotation profile)  $v_c$. With respect to the comoving frame of reference  (\ref{eq:tetrad1}) - (\ref{eq:tetrad2}), the 4-velocity of the particles ${\bf u}=d{\bf x} /ds$ has components
\begin{equation}
u^{(a)} = e^{(a)}_{\ \ \ b} u^b,
\end{equation}
while  the  3-velocity 
\begin{equation}
v^{(i)} = c \frac { u^{(i)} } { u^{(0)} }  =  c  \frac { e^{(i)}_{\ \  \ a} u^a }{  e^{(0)}_
{ \ \ \ b}  u^b  }. 
\end{equation}

For circular, equatorial  orbits   ${\bf u} =u^0(1,0,0,\omega/c)$, where  $\omega=\frac{d\varphi}{d t}=cu^3/u^0$ is  the  angular speed of the test particles, and 
 $v^{ (\varphi)}$ is the only nonvanishing velocity  component, and is given by
\begin{equation}
 [v^{ (\varphi)}]^2 = v_c^2= - \frac{ g_{33} }{ g_{00} } \omega ^2 . \label{eq:vc2}
\end{equation}
This quantity  represents   the circular speed   (rotation curves)   of the particles measured by an inertial
observer far from the source, and hence the equality. The angular speed  $\omega$ can be calculated  considering  the radial motion of the particles along geodesics.
The Lagrangian density for a massive test particle is defined as 
\begin{equation}
2{\cal L} = g_{ab} \dot x^a \dot x^b =  e^{2\nu} c^2 \dot t^2 -  e^{2\lambda}  (\dot r^2 + r^2 \dot  \theta ^2 + r^2
\sin^2 \theta  \dot \varphi ^2),
\end{equation}
where  the overdot denotes derivative with respect to the affine parameter $s$ and   Lagrange's equations read
\begin{equation}
\frac{d}{ds} \left ( \frac{ \partial {\cal L}} {\partial \dot x^a} \right )
- \frac{ \partial {\cal L}} {\partial x^a} = 0. 
\end{equation}

For circular orbits in the equatorial plane 
$\dot r =\dot \theta = 0$, and the equation of motion for the radial coordinate gives
\begin{equation}
\omega ^2 = - c^2 \frac{g_{0 0, r}}{g_{33, r }}.
\end{equation} 

The tangential velocity  is then 
\begin{equation}
v_c^2  =
c^2 \frac {r \nu_{,r}}{1+r \lambda_{,r}}. 
\end{equation}

The stability of orbits against radial perturbations can be analyzed  using an extension of the Rayleigh criteria for stability of a fluid at rest in a
gravitational field \cite{RAYL,LAND6}. The stability condition is  \cite{Let-estab}
\begin{equation}
\frac{d(h^2)}{dr} \ > \ 0 ,
\end{equation}
where $h$  is the specific
angular momentum of a particle rotating at a radius $r$, defined as $h =p_\varphi=- c \frac{ \partial {\cal L}} {\partial \dot \varphi} = -  g_{33} u^0 \omega $.
$u^0$ is obtained normalizing  $u^a$, that is requiring  $g_{ab}u^au^b=1$, so that
\begin{equation}
 (u^0)^2 =  \frac {1}{g_{00} +g_{33} \omega^2/c^2} \label{eq:u0}.
\end{equation}
 For the equatorial plane $\theta =\pi /2$, we have
\begin{equation}
h^2 =  \frac { e^{2 \lambda} r^2 v_c^2 }  { 1- \frac{v_c^2}{c^2}  }   .
\label{eq:moman}
\end{equation}

\section{ Harmonic oscillator type spheres }
In Newtonian gravity the gravitational potential of  a  sphere of radius $a$ and constant  mass density  $\rho_N$  is 
\begin{equation}
  \Phi=\begin{cases}
    -2 \pi G \rho_N (a^2 -  \frac 1 3 r^2)  &  (r<a), \\
     - \frac{4 \pi G \rho_N a^3}{3 r} &  (r>a).
  \end{cases}
\end{equation}
For $r<a$ the potential corresponds to a  harmonic oscillator potential which  has  been  used to model extended dark matter haloes  with harmonic
core \cite{Wilkinson}. The circular speed is
\begin{equation}
v_{cN}= \sqrt{ \frac{4 \pi G \rho_N}{3}} r. 
\end{equation}

Solving the condition of pressure isotropy (\ref{eq:iso}),  the interior solution is
\begin{subequations}\begin{eqnarray}
e^{ \nu} &= & \left [  C_1 \left ( \frac{r^2}{a^2} - \frac{3(b+2)}{b} \right )^4 
+  C_2  \left (\frac{r^2}{a^2} - \frac{3(b+2)}{b} \right )^{-3}   \right ] \left [ 
1 + \frac b 2 \left (1- \frac 1 3 \frac{r^2}{a^2} \right )   \right]^2, \\
 e^{ \lambda} &= &  \left [ 
1 + \frac b 2 \left (1- \frac 1 3 \frac{r^2}{a^2} \right )   \right]^2,
\end{eqnarray} \end{subequations} 
where $b = 2 \pi G \rho_N a^2 / c^2$ is the parameter that measures the strength of the gravitational field, and $C_1$ and
$C_2$ are  constants of integration which obtain by demanding the continuity  of the first and the second
fundamental forms (Darmois conditions) at the boundary $r=a$ \cite{Darmois,Bonnor-junt}
between the above interior space-time and the  exterior Schwarzschild metric which in isotropic coordinates reads
\begin{subequations}\begin{eqnarray}
e^{ \nu_{Sch}} &= & \frac {1- \frac{ab}{3r}}{1+ \frac{ab}{3r}}, \\
 e^{ \lambda_{Sch}} &= &  \left (
1+ \frac{ab}{3r}
\right)^2.
\end{eqnarray} \end{subequations} 
Then it follows 
\begin{subequations}\begin{eqnarray} 
 C_1 &=& - \frac{9b^4}{112(b+3)^6},  \\
 C_2   &=&  \frac{432 (b-4 )}{7 b^3 }.
\end{eqnarray} \end{subequations} 

\begin{figure}
$$
\begin{array}{cc}
\tilde \rho   & \tilde p  \\
\includegraphics[width=0.3\textwidth]{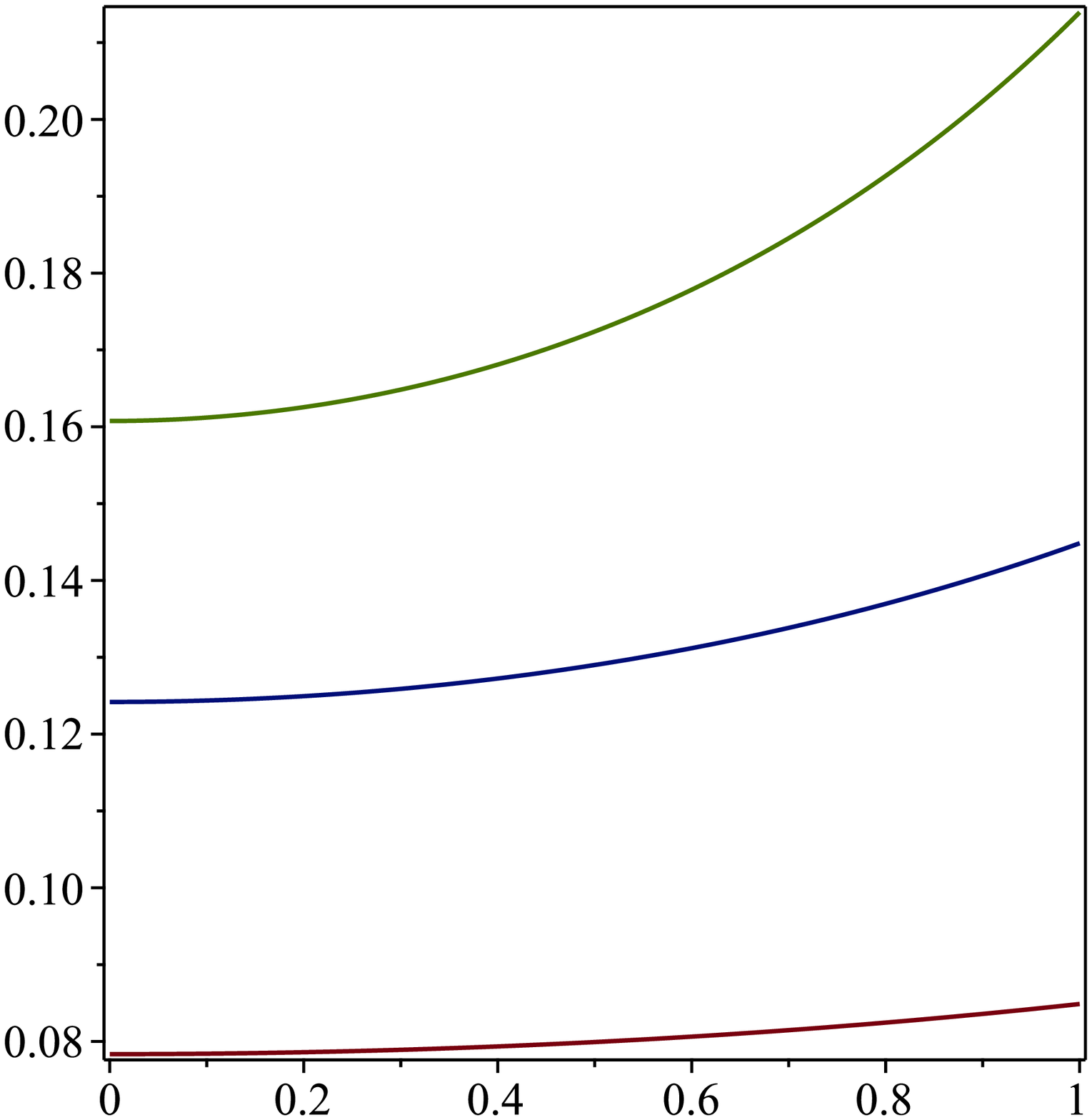} &
\includegraphics[width=0.3\textwidth]{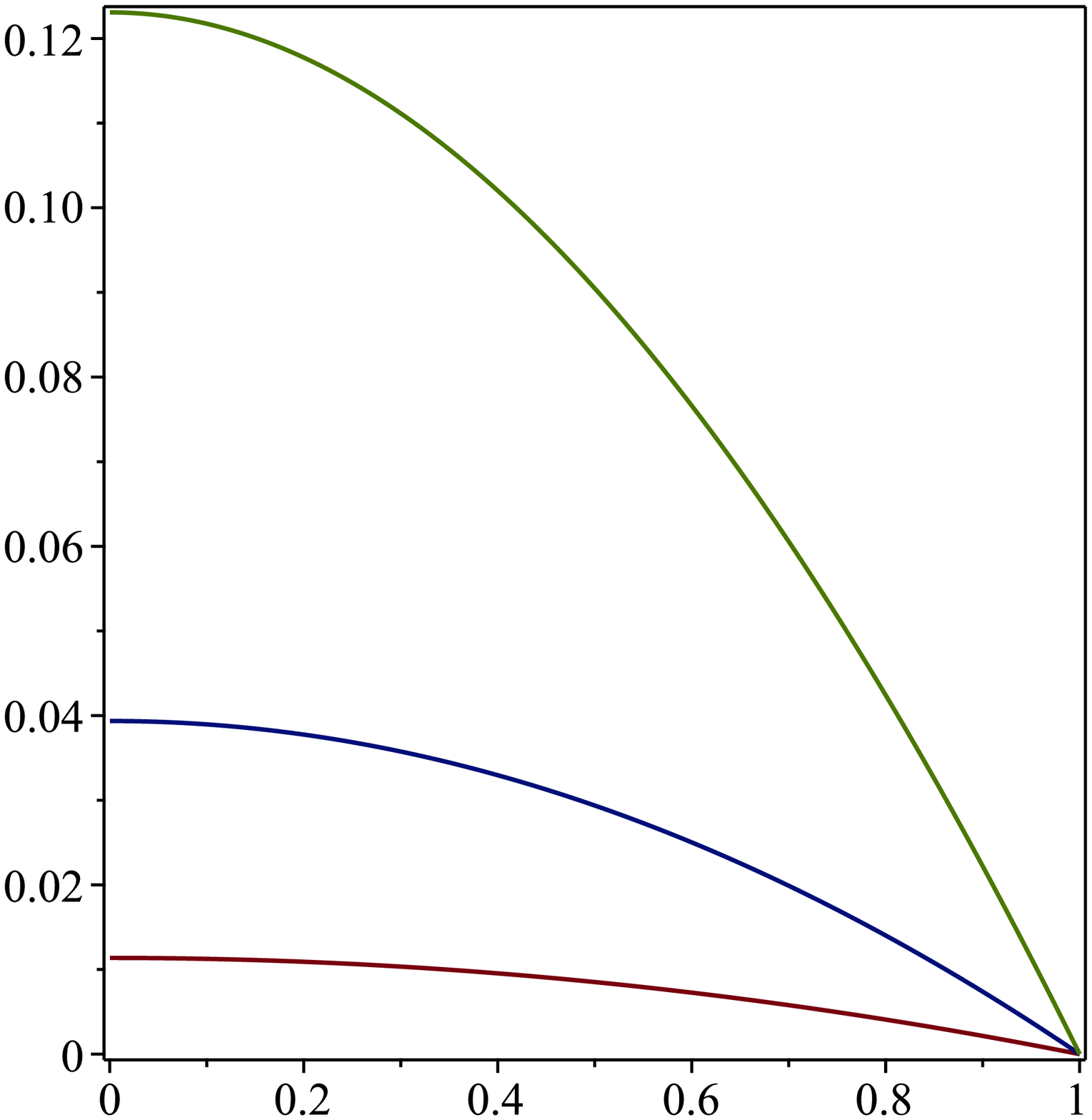}  \\
\tilde r & \tilde r \\
&  \\
(a)    &  (b)  \\
& \\
\tilde v_c^2  &  \tilde h^2  \\
\includegraphics[width=0.3\textwidth]{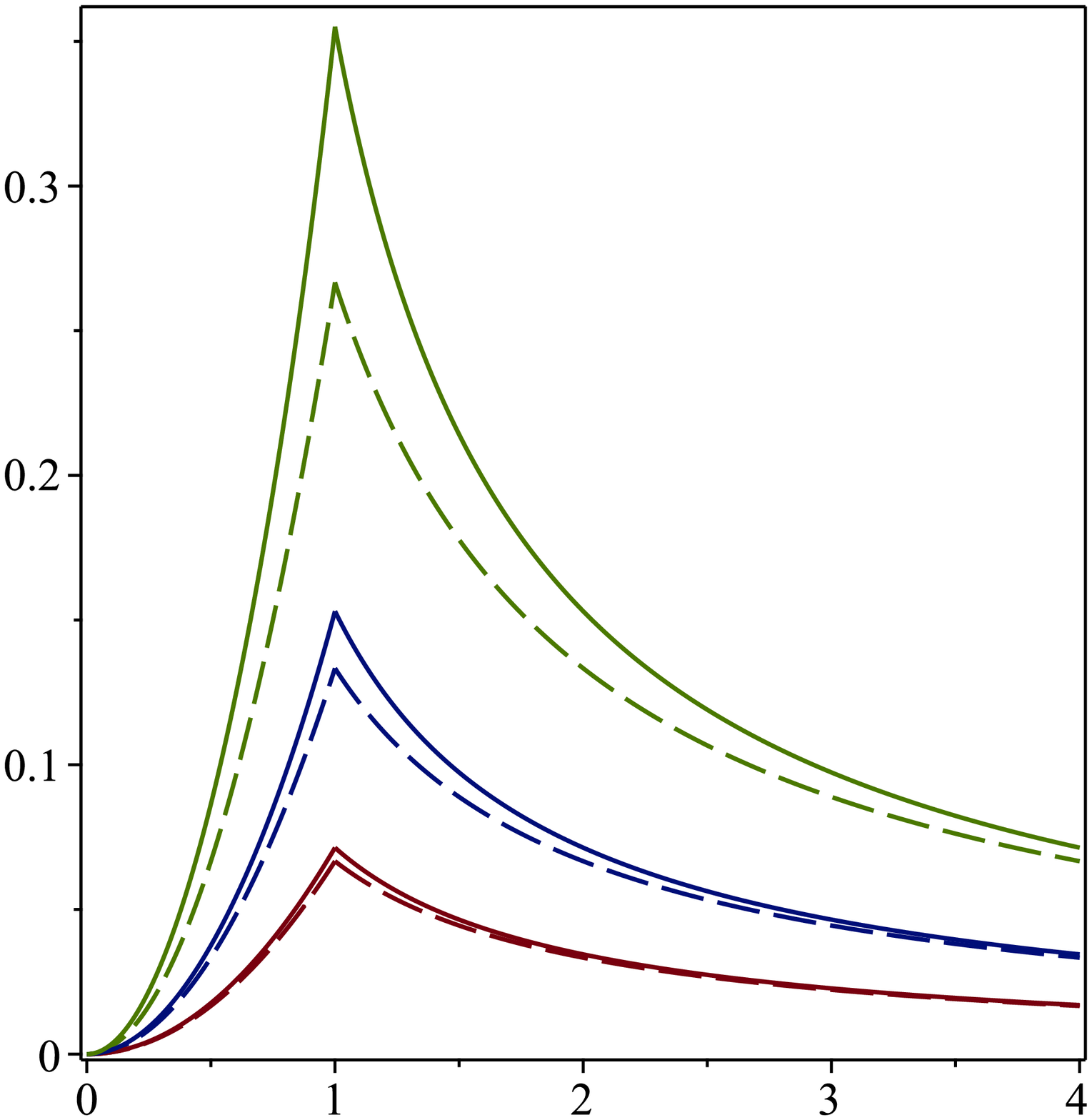} &
\includegraphics[width=0.3\textwidth]{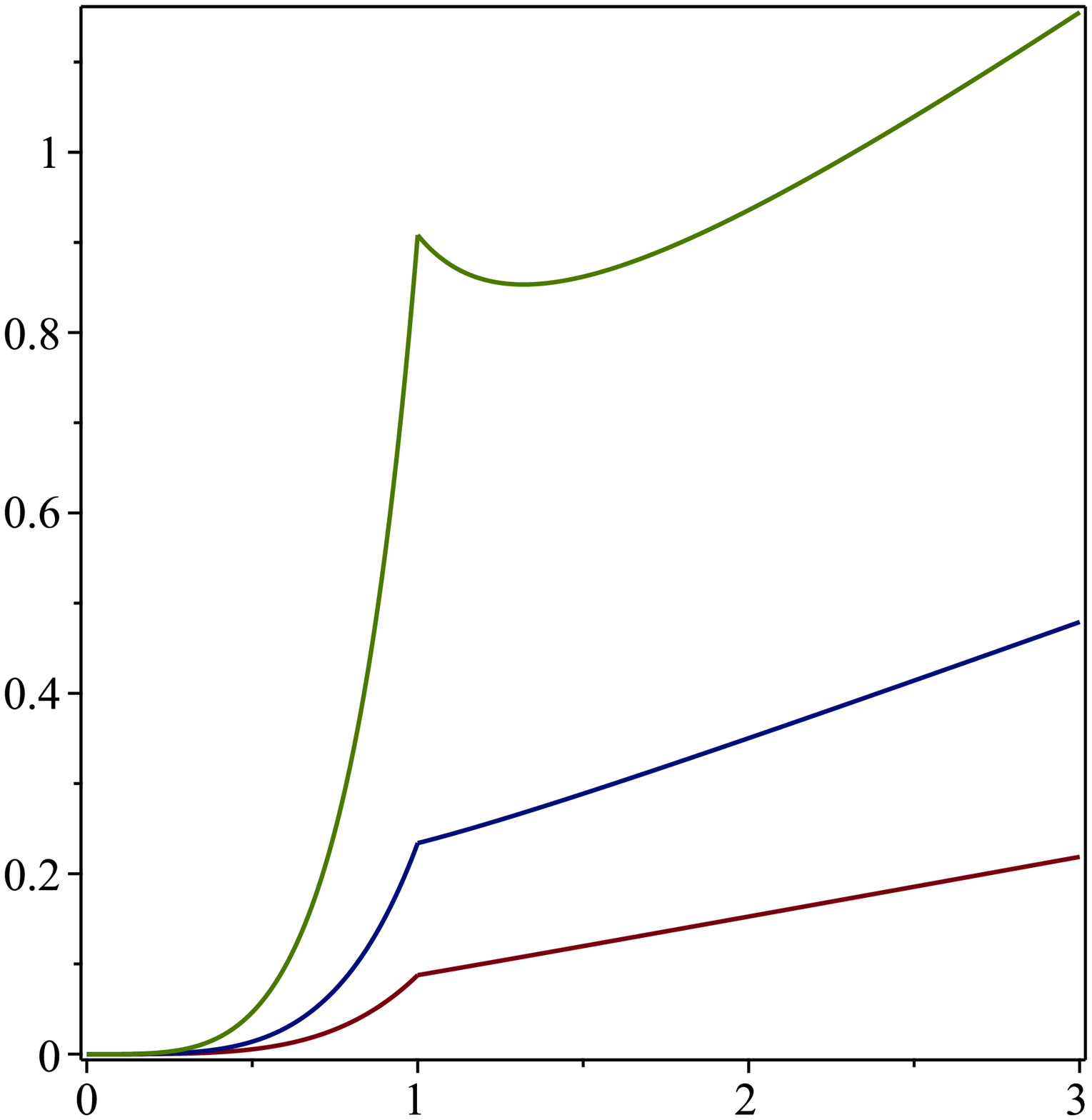}  \\
 \tilde r & \tilde r   \\
 &   \\
(c)    &  (d)
\end{array}
$$	
\caption{ For the relativistic analogue of a homogeneous sphere  we plot,     as functions of $\tilde r$,  $(a)$ the energy density  $\tilde \rho$, $(b)$ the isotropic pressure $\tilde p$    with   parameter  $ b = 0.1$  (bottom curves),  $0.2$,  $0.4$ (top curves),  $(c)$ the relativistic circular speed $\tilde v_c^2$
(solid curves),   the Newtonian rotation curves  $\tilde v_{cN}^2$ (dashed curves) with  $ b = 0.1$  (bottom curves),  $0.2$,  $0.4$ (top curves), and
   $(d)$  the specific angular momentum $\tilde h^2$   for the same value of the  parameter $ b = 0.1$  (bottom curve),  $0.2$  ,  $0.4$ (top curve).    }
\label{fig:fig1}
\end{figure}

The main relativistic  physical quantities associated with these matter distributions  are
\begin{subequations}\begin{eqnarray}
\rho & = & \frac{\rho_N}{\left[  1+ \frac b 2 (1 - \frac 13 \tilde r ^2 )  \right]^5}, \\
p & = & - \frac {64 b c^4}{3 \pi G a^2} \frac { \left [ \left[  \left( 1-\frac 1 3 \tilde r ^2  \right) b+2 \right] ^
{6} \left[  \left( 1 - \frac 3 2 \tilde r ^2  \right) b+2  \right]  + \frac {32} {729} \left( b-4 \right)  \left( b+3 \right) ^{6}  \right ] }
{ \left [ \left[  \left( 1 - \frac 1 3 \tilde r ^2 \right) b+2 \right] ^
{7}  + \frac {256} {729} \left( b-4 \right)  \left( b+3 \right) ^{6} \right ] 
 \left[  \left( 1-\frac 1 3 \tilde r ^2 \right) b+2 \right] ^{5} },  \\
 v_c^2 & = & 4 b c^2 \tilde r ^2 \frac{ \left [-  \left[  \left( 1 - \frac 1 3 \tilde r ^2 \right) b+2 \right] ^
{7}  + \frac {128} {2187} \left( b-4 \right)  \left( b+3 \right) ^{6}   \right ]}
{\left [ \left[  \left( 1 - \frac 1 3 \tilde r ^2 \right) b+2 \right] ^
{7}  + \frac {256} {729} \left( b-4 \right)  \left( b+3 \right) ^{6} \right ] \left[  \left( 1-\frac 5 3 \tilde r ^2 \right) b+2 \right] }, \\
h^2 & = & \frac { a^2 b c^2}4 \tilde r ^4   \frac { \left [  -  \left[  \left( 1 - \frac 1 3 \tilde r ^2 \right) b+2 \right] ^
{7}  + \frac {128} {2187} \left( b-4 \right)  \left( b+3 \right) ^{6}   \right ]  \left [ \left( 1 - \frac 1 3 \tilde r ^2 \right) b+2 \right]^4 }
{\left[ \left( 1 - \frac 1 3 \tilde r ^2 \right) b+2 \right] ^
{7} \left[  \left( 1 + \frac 7 3 \tilde r ^2 \right) b+2 \right]  + \frac {256} {729} \left( b-4 \right)  \left( b+3 \right) ^{6}   \left[  \left( 1 - \frac 7 3 \tilde r ^2 \right) b+2 \right]   
 } ,
\end{eqnarray} \end{subequations}
where $\tilde r = r /a$.
Since $r^2/a^2 \leq 1$ the energy density   is always  positive  in accordance  with the weak energy condition. The other physical quantities will be analyzed using a graphical method. 
In  figure \ref{fig:fig1} we plot the  energy density  $\tilde \rho = 2\pi G a^2 \rho /c^2$, the isotropic pressure $\tilde p = (8 \pi G /c^4) p$, the circular speed $\tilde v_c^2 = v_c^2/c^2$,  the Newtonian rotation curves  $\tilde v_{cN}^2= v_{cN}^2/c^2 $   and the specific angular momentum $\tilde h^2= h^2/(c^2 a^2)$ for the relativistic analogue of a homogeneous sphere  with   gravitational parameter  $ b = 0.1$,    $0.2$,   $0.4$,    as functions of $\tilde r $. We see that  the energy density becomes constant as the gravitational field decreases, and  the stresses are positive (pressure) and vanish at the boundary $r=a$.   We also observer that the relativistic effects 
increase everywhere  the speed of particles  and they  become  more important  as we move away from the central region. Moreover,  we find that the increase in the  gravitational field
can make  the orbits of particles  unstable against radial perturbations. Indeed,  the solution  with parameter $b=0.4 $ presents  a  region of instability of the orbits near  the boundary.  For these values of parameters the energy conditions are all satisfied. 

\section{Plummer type  spheres  }

A simple Newtonian potential-density pair is  Plummer's model \cite{Plummer}
\begin{subequations}\begin{eqnarray}
\Phi  &=&  - \frac{GM}{\sqrt{r^2 + a^2}}, \\
\rho_N &=& \left (  \frac {3M}{4 \pi a^3}   \right) 
\left( 1 + \frac {r^2}{a^2}  \right )^{-\frac{5}{2}}, 
\end{eqnarray} \end{subequations} 
where $a$ is a non-zero constant with the dimension of length.
 Plummer's  sphere  is used to model   star clusters,
 the  central  spherical  nucleus of spiral galaxies \cite{M-N1,M-N2} and also as models of dark matter haloes \cite{Bajkova2}. The circular speed is
\begin{equation}
v_{cN}^2= \frac{ GMr^2 }{\left (r^2 + a^2 \right )^{3/2}}.
\end{equation}

The relativistic version of this model  is a particular case of  Buchdahl's solutions \cite{BUCH}. Therefore, the metric functions are 
\begin{subequations}\begin{eqnarray}
e^{2 \nu} &= & \left( \frac{   1-\frac{GM}{2ac^2 \sqrt { 1 + \frac{r^2}{a^2} }} }{ 1+\frac{GM}{2a c^2  \sqrt {1 + \frac {r^2} { a^2} }} }  \right )^2,  \\
e^{2 \lambda} &=& \left( 1+\frac{GM}{2ac^2 \sqrt {1 + \frac{r^2}{a^2} } }   \right)  ^{4}. 
\end{eqnarray} \end{subequations} 

The main relativistic  physical quantities associated with these structures  are
\begin{subequations}\begin{eqnarray}
\tilde \rho &=& \frac{3 b}{2 \pi  \left (  \sqrt{1 +\tilde r^2 } + b  \right )^5} , \\
&& \\
 \tilde p&=& \frac{b} {4 \pi  \left ( \sqrt{1 + \tilde r^2 } -  b  \right )   \left ( \sqrt{1 + \tilde r^2 } + b  \right )  ^5  },
 \\
&&\\
\tilde v_c^2 &=&   \frac{ 2b \tilde  r^2 \sqrt{1+ \tilde r^2} }{ \left (   
\sqrt{1+ \tilde r^2}  - b \right)  \left[ (1+ \tilde r^2)^{3/2} + b (1-\tilde r^2)   \right]},\\
& & \\
\tilde h^2  &=&  \frac {2b \tilde r^4 \left( \sqrt{1+ \tilde r^2}  + b \right )^4} { 
 \left( 1 + {\tilde r}^{2} \right) ^{3/2} \left[ -4 b \,{\tilde r}^{2}\sqrt {1 + {\tilde r}^{2} }+
{b}^{2} \left( {\tilde r}^{2}-1 \right) + \left( 1 + {\tilde r}^{2} \right) ^{2} \right]  },
\end{eqnarray} \end{subequations} 
where $\tilde \rho = \frac{ G a^2}{c^2} \rho$, $\tilde p = \frac{a^3}{c^2 M} p$,  $\tilde v_c^2 = v_c^2 /c^2 $, $\tilde h^2 = h^2 /c^2 $,  $\tilde r = r /a$ and $b=GM/(2ac^2)$.   The energy density always is positive in agreement   with the weak energy condition and for $b<1$  we have  positive stresses, i.e. pressure.

\begin{figure}
$$
\begin{array}{cc}
\tilde \rho   & \tilde \rho_{N}  \\
\includegraphics[width=0.3\textwidth]{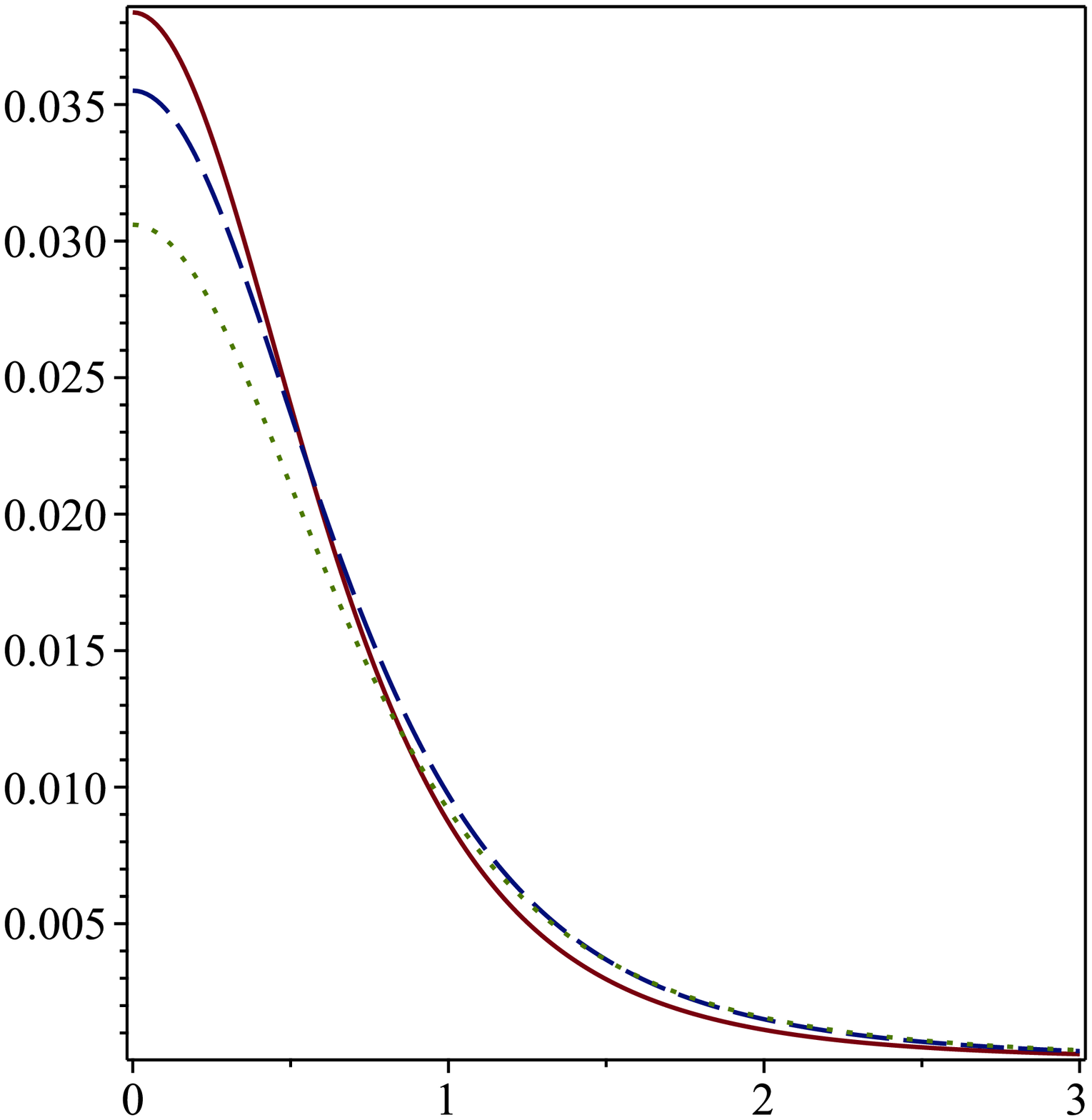} &
\includegraphics[width=0.3\textwidth]{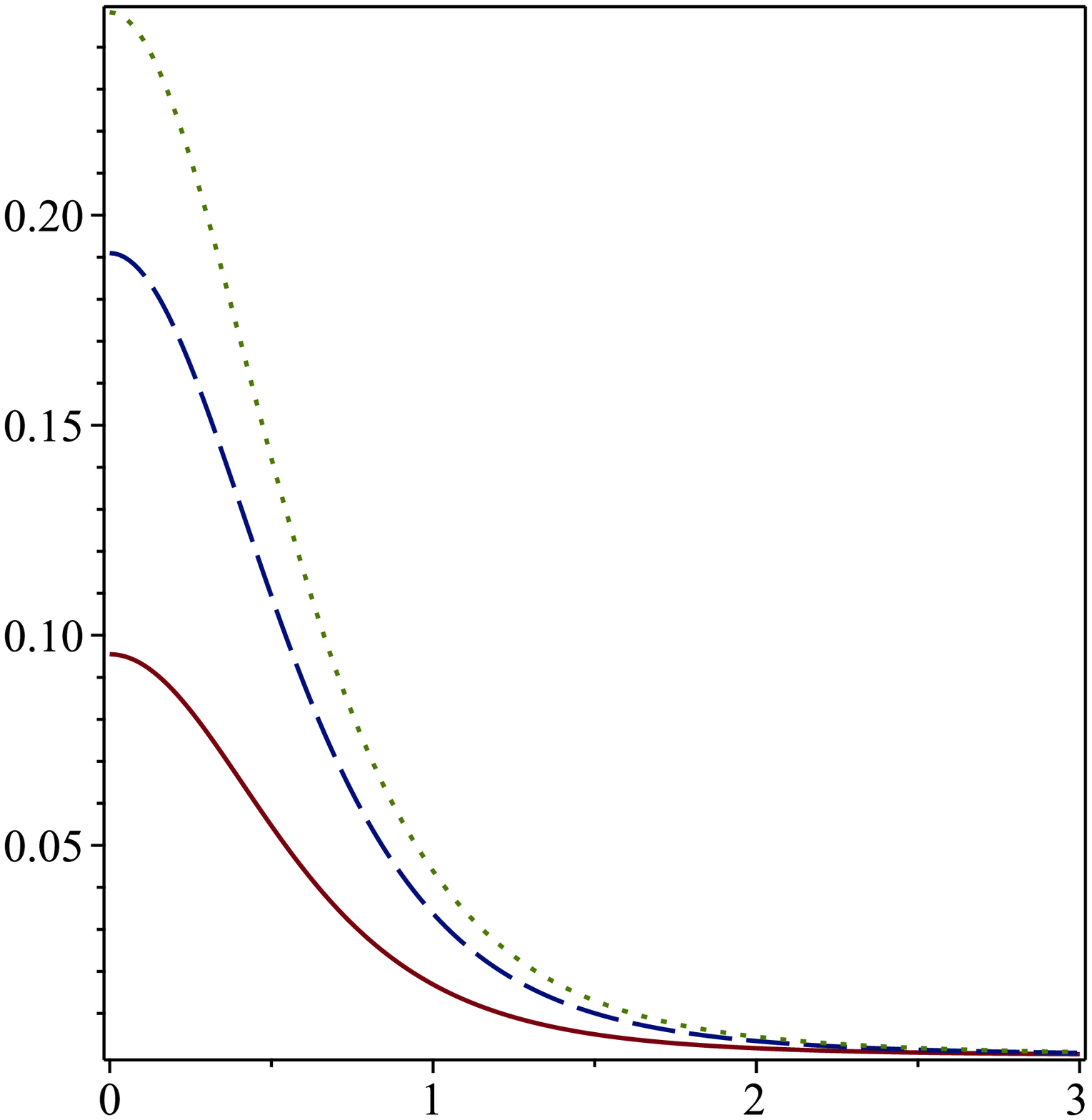}  \\
\tilde r & \tilde r \\
&  \\
(a)    &  (b)  \\
& \\
\tilde p  &   \\
\includegraphics[width=0.3\textwidth]{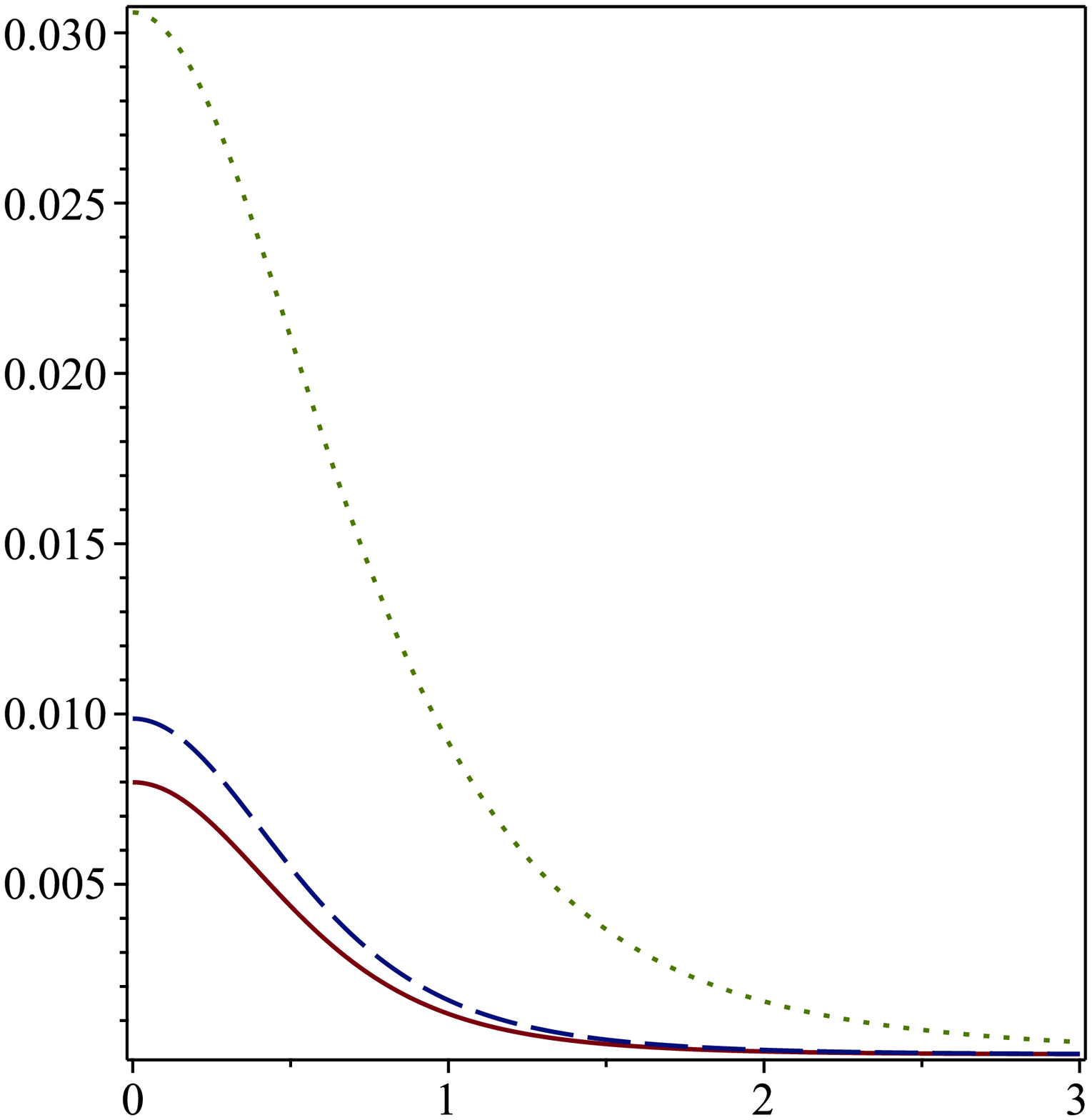} &
  \\
 \tilde r &  \\
 &   \\
(c)    & 
\end{array}
$$	
\caption{  The  relativistic and Newtonian density  profiles $(a)$ $\tilde \rho$,  $(b)$ $\tilde \rho_N$ and  $(c)$ the isotropic pressure $\tilde p$  for  Plummer type spheres   with parameter  $ b = 0.2$  (solid curves),  $0.4$  (dashed  curves),  $0.52$ (dotted curves),    as functions of $\tilde r$.  }
\label{fig:fig2}
\end{figure}

In figure \ref{fig:fig2} we graph, as  functions of $\tilde r$, the  relativistic and Newtonian density  profiles  $\tilde \rho$ and 
$\tilde \rho_N = \frac{ G a^2}{c^2}\rho_N$,  and the isotropic pressure $\tilde p$ for  Plummer type spheres   with gravitational parameter  $ b = 0.2$,  $0.4$,  $0.52$. 
We find  that the energy density presents a maximum at $r = 0$ and then
decreases rapidly with $r$. We also find that when the gravitational field  is increased   the relativistic  density profile  always decreases at the
central region of the matter distribution  but then increases.
In turn, the relativistic corrections decrease  everywhere  the energy density.  

\begin{figure}
$$
\begin{array}{cc}
 \includegraphics[width=0.3\textwidth]{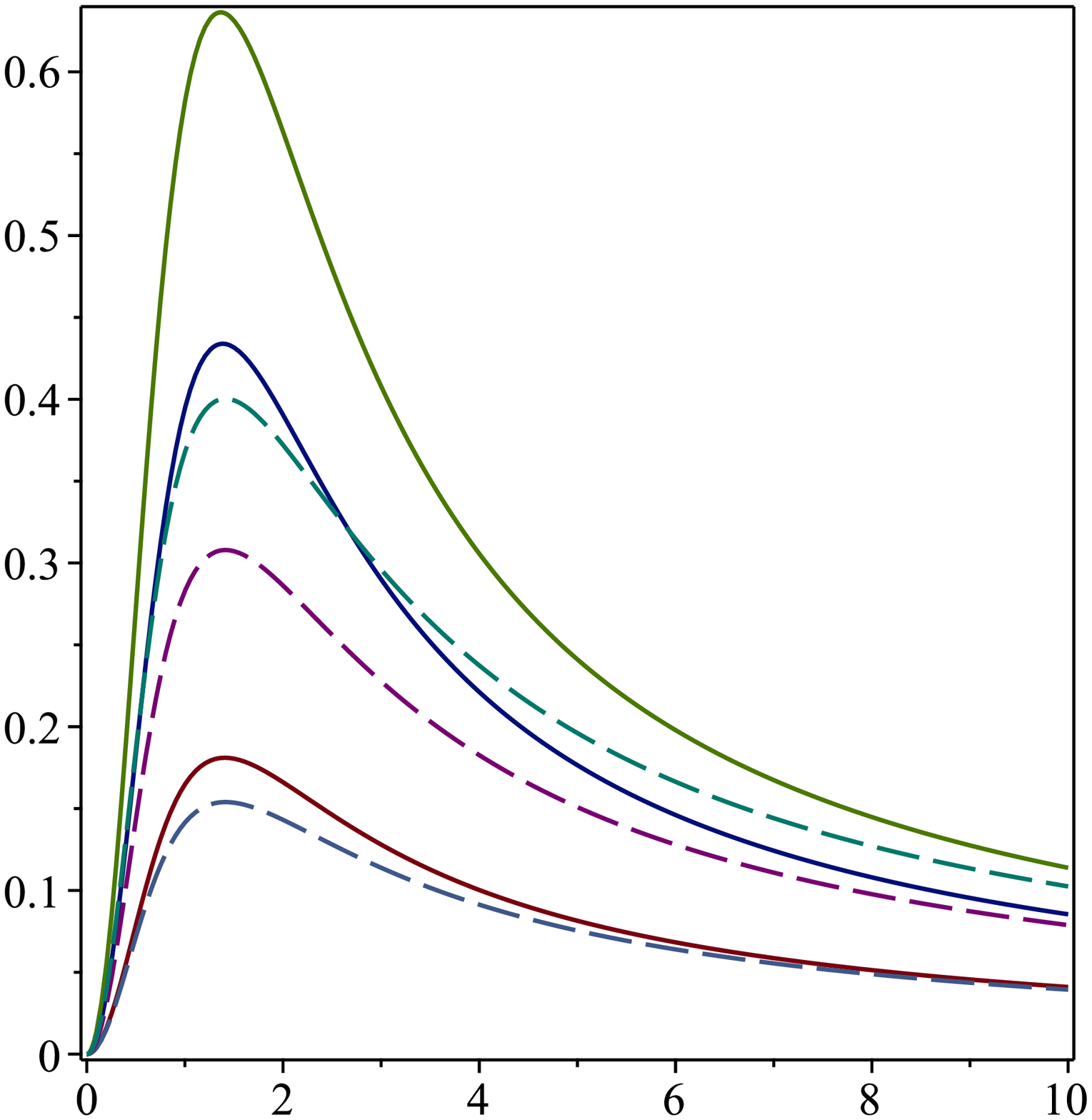} &
\includegraphics[width=0.3\textwidth]{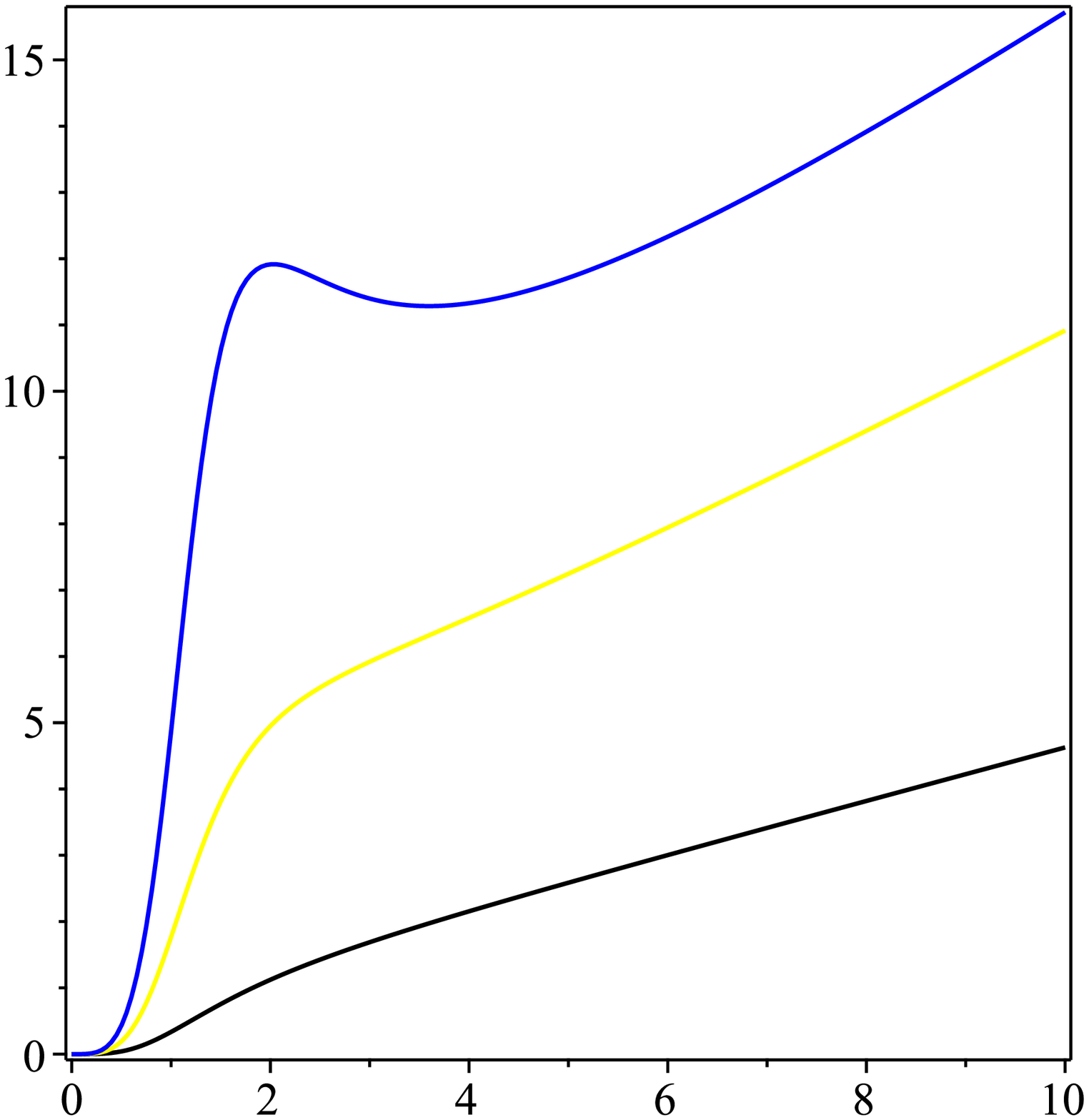}  \\
 \tilde r & \tilde r   \\
 &   \\
(a)    &  (b)  
\end{array}
$$	
\caption{ $(a)$ The  relativistic and Newtonian  circular  speed   $\tilde v_c^2$  (solid curves) and $\tilde v_{cN}^2$  (dashed curves), and  $(b)$ the specific angular momentum $\tilde h^2$  for  Plummer type spheres   with parameters  
$ b = 0.2$  (bottom curves),  
$0.4$,  $0.52$ (top curves),    as functions of $\tilde r$.     }
\label{fig:fig3}
\end{figure}

 In figure \ref{fig:fig3} we plot the relativistic and Newtonian rotation curves   $\tilde v_c^2$ and  $\tilde v_{cN}^2 = v_{cN}^2 /c^2 $,   and   the specific angular momentum $\tilde h^2$, also as  functions of $\tilde r$.
We see that  both the gravitational field and  relativistic effects   increase the 
circular speed of particles and also that such corrections are more important
in the regions around its maximum value and, as expected, for velocities comparable
to the speed of light. We  also find that the increase in the  gravitational field
can make  the orbits of particles  unstable against radial perturbations. Thus,  the solution with parameter $b=0.52$ presents a region of instability
of the orbits.  For this value of the parameter the dominant energy condition is also  satisfied. 

\section{Logarithmic potential type  spheres}
%logarithmic potential type dark-halo models
Dark matter haloes  can be  modeled in Newtonian theory with a  logarithmic potential of the  form \cite{Binney}
\begin{equation}
\Phi = \frac 12 v_0^2  \ln (\frac{r^2 + a^2}{b}),
\end{equation}
where  $a$ and $b$ are constants, and $v_0$ is  circular speed at large radii,  also a constant. When $a=0$ this potential is  often referred to as the singular isothermal sphere.   The mass density distribution is
\begin{equation}
\rho_N = \frac {v_0^2 (r^2 + 3 a^2)}{4 \pi G (r^2 +  a^2)^2} , 
\end{equation}
and the circular speed at radius $r$ is 
\begin{equation}
v_{cN} = \frac{v_0 r}{\sqrt{r^2 + a^2}}. 
\end{equation}
This potential yields an asymptotically flat rotation curve. 

\begin{figure}
$$
\begin{array}{cc}
\includegraphics[width=0.3\textwidth]{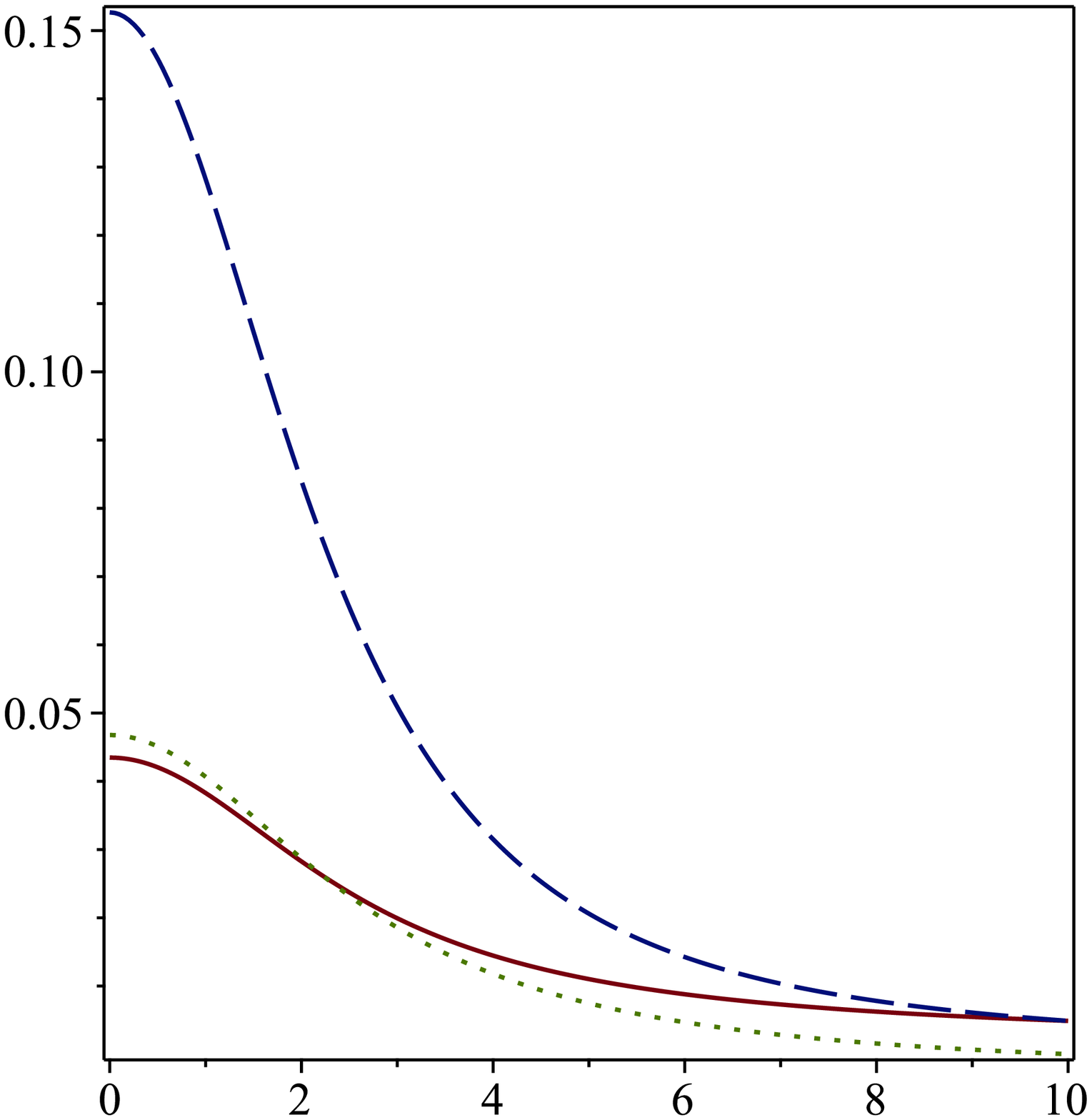} &
\includegraphics[width=0.3\textwidth]{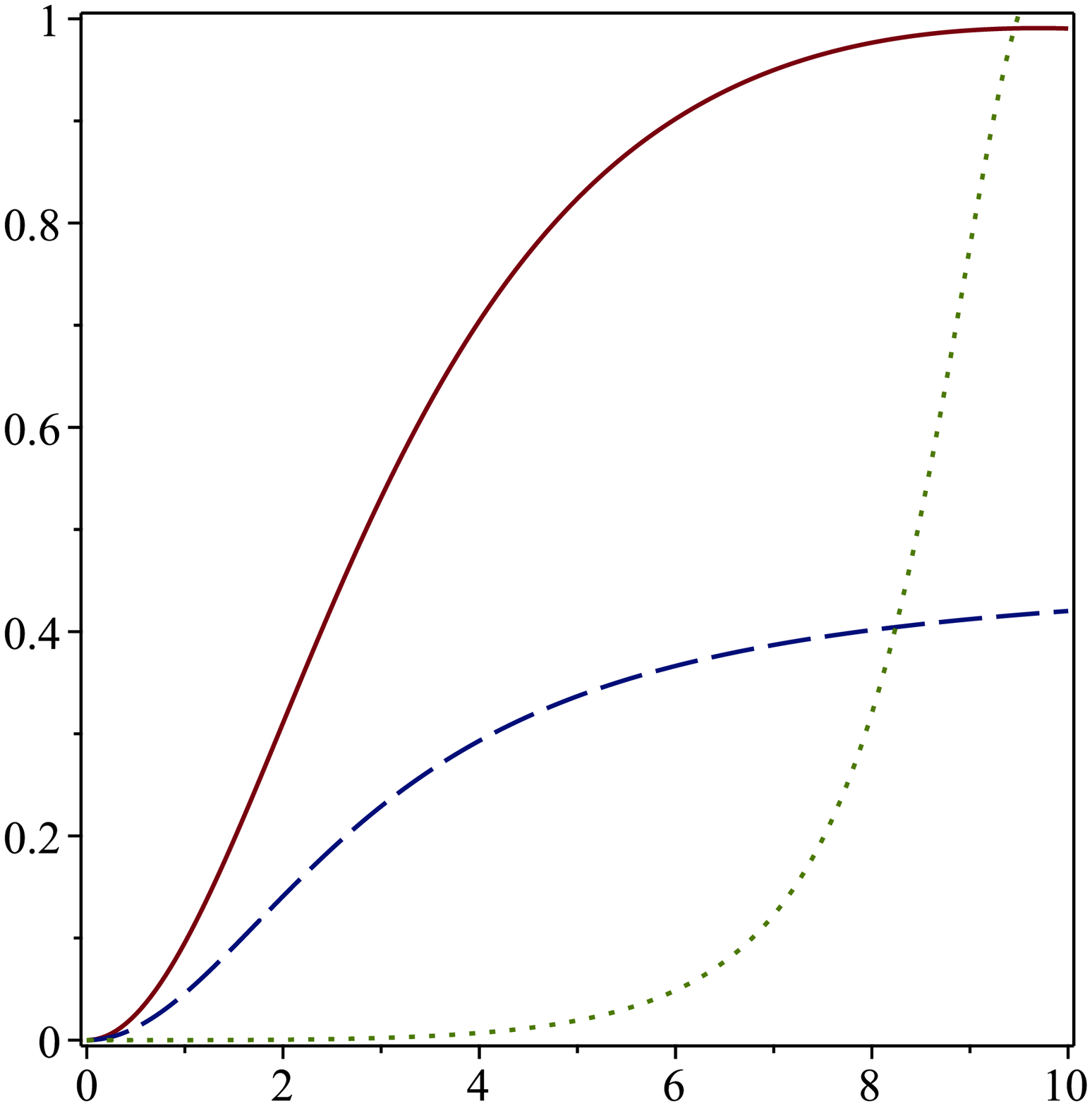}  \\
 r &  r \\
&  \\
(a)    &  (b) 
\end{array}
$$	
\caption{ $(a)$ The relativistic energy density  $\tilde \rho$ (solid curve), Newtonian density distribution $\tilde \rho_N$ (dashed curve),  the isotropic pressure $\tilde p$ (dotted curve), $(b)$ the relativistic circular speed $\tilde v_c^2$ (solid curve), Newtonian rotation curve $\tilde v_{cN}^2$ (dashed curve)  and the specific angular momentum $\tilde h^2\times10^{-4}$   (dotted curve),   for  
logarithmic potential type  dark matter haloes   with parameters  $ \tilde v_0^2 = 0.458$, $a=3$     
and $r_c=10$, as functions of $ r$.   }
\label{fig:fig4}
\end{figure}

Solving (\ref{eq:iso})  a particular solution is
\begin{subequations}\begin{eqnarray}
e^{ \nu} &= &C \left( {r}^{2} + {a}^{2} \right)  \left[ 1- \frac 14 
\, \frac{v_0^2}{c^2} \ln  \left( {
\frac { {r}^{2} + {a}^{2} }{b}} \right)  \right] ^{6} , \\
e^{ \lambda} &= & \left[ 1- \frac 14 
\,  \frac{v_0^2}{c^2} \ln  \left( {
\frac { {r}^{2} + {a}^{2} }{b}} \right)  \right] ^{2},
\end{eqnarray} \end{subequations} 
where $C$ is an  integration constant. 
The general solution is complicated and is not  presented since it introduces unphysical properties to the solution.
The main relativistic  physical quantities associated with the   distribution of matter  are 
\begin{subequations}\begin{eqnarray}
\tilde \rho & = & {\frac {\tilde v_0^2 \left( {r}^{2} + 3 {a}^{2} \right) }{ \left( {r}^{2} + {a}^{2}
 \right) ^{2}} \left[ 1-\frac 14 \, \tilde v_0^2\ln  \left( {\frac {{{r}^{2} + a}^{2}}{b}}
 \right)  \right] ^{-5}} ,  \label{eq:rho} \\
\tilde  p & = &  \frac { 2048 \, \tilde v_0^2  \left( (7 \tilde v_0^2 -4)r^2  + (5 \tilde v_0^2 -4) a^2  \right) \ln  \left( {\frac {{r}^{2} + {a}^{2} }{b}} \right)  +\left( 28672\, \tilde {v_0}^{4}-49152\,\tilde v_0^2 +16384 \right) {r}^{2}-32768\,{a}^{2}  
 \left( \tilde v_0^2 - \frac 12 \right)  } {4096\left( {r}^{2} + {a}^{2}
 \right) ^{2} \left[ 1-\frac 14 \, \tilde v_0^2\ln  \left( {\frac {{{r}^{2} + a}^{2}}{b}}
 \right)  \right] ^{6}} ,  \nonumber \\
& &  \\
 \tilde v_c^2&=& \frac { 2\,{r}^{2} \left[ \tilde v_0^2 \ln  \left( {\frac {{r}^{2} +{a}^{2}}{b}} \right) +6
\,\tilde v_0^2-4 \right]  }{ \tilde v_0^2 \left( {r}^{2} + {a}^{2} \right) \ln  \left( {\frac {{r}^{2} + {a}^{2}}{
b}} \right) +  4(\tilde v_0^2-1){r}^{2}  -4\,{a}^{2}
}, \\
\tilde h^2 & =& \frac{ {r}^{4} \left[ \tilde  v_0^2  \ln  \left( {\frac {{r}^{2}+ a^2}{b}} \right) -4 \right] ^
{4} \left[ \tilde v_0^2 \ln  \left( {\frac {{r}^{2} + {a}^{2}}{b}} \right) +6
\,\tilde v_0^2-4 \right] } { 128 \left [ \left( {a}^{2}-{r}^{2} \right) \tilde v_0^2 \ln  \left( {\frac {
{r}^{2} +{a}^{2}}{b}} \right) +4{r}^{2}(1-2 \tilde v_0^2)  - 4\,{a}^{2} \right ]  }  ,
\end{eqnarray} \end{subequations} 
where $\tilde \rho = \frac{4 \pi G}{c^2} \rho$,  $\tilde p = \frac{8 \pi G}{c^4} p$,    $\tilde v_c^2 = v_c^2/c^2$,     $\tilde v_0^2 = v_0^2/c^2$  and $\tilde h^2= h^2/c^2 $.  In order for   the weak energy condition $\rho \geq 0$ to be satisfied 
a cut-off radius $r_c$ must be imposed 
such  that  $r_c^2+a^2=b$. 
Again,  the other relationships   will be analyzed  using  a graphic method.  In  figure \ref{fig:fig4} $(a)$  we graph,  as functions of $ r$, 
the relativistic energy density  $\tilde \rho$, the Newtonian density profile $\tilde \rho_N$ and  the isotropic pressure $\tilde p$   for  
dark matter haloes  constructed from a logarithmic seed potential   with parameters  $ \tilde v_0^2 = 0.458$, $a=3$ and $r_c=10$. 
 We observe that the energy density
presents a maximum at the center of the distribution of matter, and then decreases
rapidly with the radial distance  $r$ which  permits us to  define  a cut-off radius $r_c$  and so, in principle,   
consider the structure as  a compact object. We also see that the relativistic effects  decrease  the density profile
everywhere of the dark halo  and they  become  more important in the central region of the distribution.  At large $r$ both density profiles  (relativistic and Newtonian)   coincide. For these values of the parameters also   we have positive stresses (pressure).

 In figure \ref{fig:fig4} $(b)$    we show
the relativistic circular speed $\tilde v_c^2$, the Newtonian rotation curve $\tilde v_{cN}^2=v_{cN}^2/c^2$  and the specific angular momentum $\tilde h^2 \times 10^{-4}$ 
for the same values of parameters, also as function of $r$.
In  contrast to  the  energy density, we  find    that  the relativistic effects  increase everywhere  the circular speed of the particles   and  they  become more  important  as we move away from the central region. 
We also see that  relativistic rotation curve is  flattened after
a certain value of  $r$ as observational data indicate. 
The speed of the particles always is less than light speed  (dominant energy condition).
We find that for these values of parameters the motion of particles is stable against radial perturbations.

\section{Conclusions}

Perfect fluid sources for  static spherically  symmetric  fields in isotropic coordinates  
 from  given solutions of    Poisson's equation   were investigated.
The   method
was  illustrated    with  three  simple examples based on the
potential-density pairs corresponding to a  harmonic oscillator, the well-known Plummer model   and a  massive spherical dark matter halo model with a logarithmic potential.  

 The geodesic circular  motion of test particles  around  such  structures  was also studied.  We found that the relativistic effects  increase everywhere  the speed of particles,
 and  are more
significant as we move away from the central region in the case of the  first and third model, in the regions close to its maximum value in the case of Plummer type fields
and, as expected, for velocities comparable
to the speed of light.  For logarithmic potential type  spheres   we also found that relativistic rotation curve is flattened after a certain value of  the radial distance as observational data indicate.

Moreover, the stability of the orbits  against radial perturbations  was  analyzed  using an extension of the Rayleigh criteria of stability of a fluid at rest in a gravitational field. In all the cases we found stable circular orbits, 
but   for   harmonic oscillator and Plummer type fields  was observed that  the increase in the  gravitational field
can make  unstable the motion of the particles. The  models considered  satisfy  all the energy conditions.

%\end{document}

%figura 1 (esfera finita)

%figura 2

%figura 3

%figura 4

\end{document}